%% file: sn-article.tex
\documentclass[pdflatex,sn-mathphys-num]{sn-jnl}

\usepackage{graphicx}%
\usepackage{multirow}%
\usepackage{amsmath,amssymb,amsfonts}%
\usepackage{amsthm}%
\usepackage{mathrsfs}%
\usepackage[title]{appendix}%
\usepackage{xcolor}%
\usepackage{textcomp}%
\usepackage{manyfoot}%
\usepackage{booktabs}%
\usepackage{algorithm}%
\usepackage{algorithmicx}%
\usepackage{algpseudocode}%
\usepackage{listings}%

\usepackage{subcaption} 
\usepackage{tabularx}
\newcolumntype{L}{>{\raggedright\arraybackslash}X} 
\newcolumntype{C}[1]{>{\centering\arraybackslash}p{#1}}
\newcolumntype{R}[1]{>{\raggedleft\arraybackslash}p{#1}}

\DeclareMathOperator*{\argmax}{argmax}

\usepackage{xcolor}

\theoremstyle{thmstyleone}%
\newtheorem{theorem}{Theorem}
\newtheorem{lemma}[theorem]{Lemma}
\theoremstyle{thmstyletwo}%

\theoremstyle{thmstylethree}%

\begin{document}

\title[Article Title]{GaloisSAT: Differentiable Boolean Satisfiability Solving via Finite Field Algebra}


\author[1]{\fnm{Curie} \sur{Kim}}
\author[1]{\fnm{Carsten} \sur{Portner}}
\author[1]{\fnm{Mingju} \sur{Liu}}
\author[2]{\fnm{Steve} \sur{Dai}}
\author[3]{\fnm{Haoxing} \sur{Ren}}
\author[3]{\fnm{Brucek} \sur{Khailany}}

\author[5]{\fnm{Alvaro} \sur{Velasquez}}
\author[6]{\fnm{Ismail} \sur{Alkhouri}}

\author*[1,4]{\fnm{Cunxi} \sur{Yu}}\email{cunxiyu@umd.edu}

\affil[1]{\orgname{University of Maryland}, \orgaddress{\city{College Park}, \state{MD}, \country{USA}}}

\affil[2]{\orgname{NVIDIA Research}, \orgaddress{\city{Santa Clara}, \state{CA}, \country{USA}}}

\affil[3]{\orgname{NVIDIA Research}, \orgaddress{\city{Austin}, \state{TX}, \country{USA}}}

\affil[4]{\orgname{NVIDIA Research}, \orgaddress{\city{College Park}, \state{MD}, \country{USA}}}

\affil[5]{\orgname{University of Colorado Boulder}, \orgaddress{\city{Boulder}, \state{CO}, \country{USA}}}

\affil[6]{\orgname{University of Michigan},\orgaddress{\city{Ann Arbor}, \state{MI}, \country{USA}}}


\abstract{Boolean satisfiability (SAT) problem, first problem proven to be non-deterministic polynomial-time (NP) complete, has become a fundamental challenge in computational complexity, with widespread applications in optimization and verification across a wide range of domains. Despite significant algorithmic advances over the past two decades, the performance of SAT solvers has improved at a limited pace. Evidently, the 2025 competition winner shows about a 2$\times$ improvement over the 2006 winner in SAT Competition performance, with almost 20 years of efforts. This paper introduces \textit{GaloisSAT}, a novel hybrid {GPU-CPU SAT solver} that integrates a differentiable SAT solving engine powered by modern machine learning infrastructure on GPUs, followed by a traditional {CDCL}-based SAT solving stage on CPUs. GaloisSAT is benchmarked against newest versions of state-of-the-art solvers, \textit{Kissat} and \textit{CaDiCaL}, using the \textit{SAT Competition 2024} benchmark suite. Results demonstrate substantial improvements in the official SAT Competition \cite{SATCompetitionWeb} metric \textbf{PAR-2}\footnote{Penalized average runtime with a timeout of 5,000 seconds and a timeout penalty factor of 2.}. Specifically, GaloisSAT achieves \textbf{8.41$\times$} speedup in the \textit{Satisfiable} category and \textbf{1.29$\times$} speedup in the \textit{Unsatisfiable} category compared to the strongest baselines.
}

\keywords{Boolean Satisfiablity, NP-Complete, Differentiable Optimization, Combinatorial Optimization, GPU Acceleration, Machine Learning}



\maketitle

\include{1-Intro}

\include{3-Methodology}
\include{4-Experiments}

\include{2-Preliminary}

\include{5-Conclusion}
\input{6-appendix}



\bibliography{bibs/cunxi,bibs/references}

\end{document}

%% file: 1-Intro.tex



\section{Introduction}\label{introduction}

The Boolean satisfiability (SAT) problem is a foundational problem of theoretical computer science. Since it was proven NP-complete by~\cite{cook2023complexity} in 1971, SAT has served as a canonical reference problem for a wide range of complex decision problems~\cite{cook2000p}. 
SAT forms the foundation of numerous applications in formal verification, electronic design automation (EDA), planning, combinatorial design, and cryptography.
A SAT formula is usually encoded in Conjunctive Normal Form (CNF) as a conjunction of clauses, where each clause is a disjunction of literals. A literal corresponds to either a Boolean variable or its negated form, with each variable taking a truth value of 1 (True) or 0 (False). The SAT problem asks whether there exists an assignment of truth values to all variables such that every clause is satisfied, meaning each clause contains at least one literal evaluated as true. If no such assignment exists, the formula is classified as unsatisfiable.


The international SAT Competition has been held annually since 2002, evaluating SAT solvers on a wide range of contributed benchmarks under a fixed time limit~\cite{SATCompetitionWeb}. It has served as a key platform in advancing SAT solving by fostering the development of new solver techniques and by continuously introducing challenging benchmarks from diverse application domains~\cite{Chen2022,Zheng2023}. Guided by the SAT competition, modern SAT solvers largely rely on the Conflict-Driven Clause Learning (CDCL) algorithm, initially introduced in the solver GRASP~\cite{marques2002grasp}, is widely used in modern SAT solving. Extending the Davis-Putnam-Logemann-Loveland (DPLL) framework~\cite{davis1962machine}, CDCL tentatively assigns variables and performs non-chronological backtracking when a conflict arises. Whenever partial assignments lead to unsatisfiability, a new clause is derived and added to the formula to prevent similar conflicts in the future. 
Despite extensive research efforts driven by the SAT Competition, the fundamentally sequential nature of CDCL makes effective parallelization difficult, limiting scalability on conventional architectures. As a result, comparatively little progress has been made toward exploiting GPUs for massive parallelization in SAT solving.


\textit{GaloisSAT} introduces a novel GPU approach that preserves the exact combinatorial logical structure in differentiable optimization fasion. Unlike previous methods that rely on massive matrix multiplications or neural networks, \textit{GaloisSAT} leverages \textbf{finite field algebraic modeling} that exactly formulates SAT problems in conjunctive normal format as a differentiable optimization problem. 
Most importantly, unlike prior incomplete approaches, our framework maintains full coverage of the search space by ensuring that all subproblems are explored, thereby guaranteeing completeness and providing valid UNSAT proofs. Moreover, \textit{GaloisSAT} demonstrates substantial performance gains over state-of-the-art solvers. When compared against the strongest baseline, \textit{Kissat}, \textit{GaloisSAT} achieves an \textbf{8.41$\times$} speedup in the \textit{Satisfiable} category and a \textbf{1.29$\times$} speedup in the \textit{Unsatisfiable} category, as measured by the PAR-2 metric. 
Notice that, our experimental evaluations show that over nearly \textbf{two decades}, the performance of the 2025 winning solver has improved only by \textbf{2$\times$} speedup compared to the 2006 winner, \textit{MiniSAT}~\cite{een2006minisat}. Together, these results establish \textit{GaloisSAT} as a novel framework that integrates precise logical modeling with high-performance computation, thereby setting a new directions for SAT solving as well as combinatorial optimizations in general.

%% file: 3-Methodology.tex
\section{Methodology}\label{methodology}

\subsection{GaloisSAT Architecture Overview}

\begin{figure}[ht]
\centering
\includegraphics[width=\textwidth]{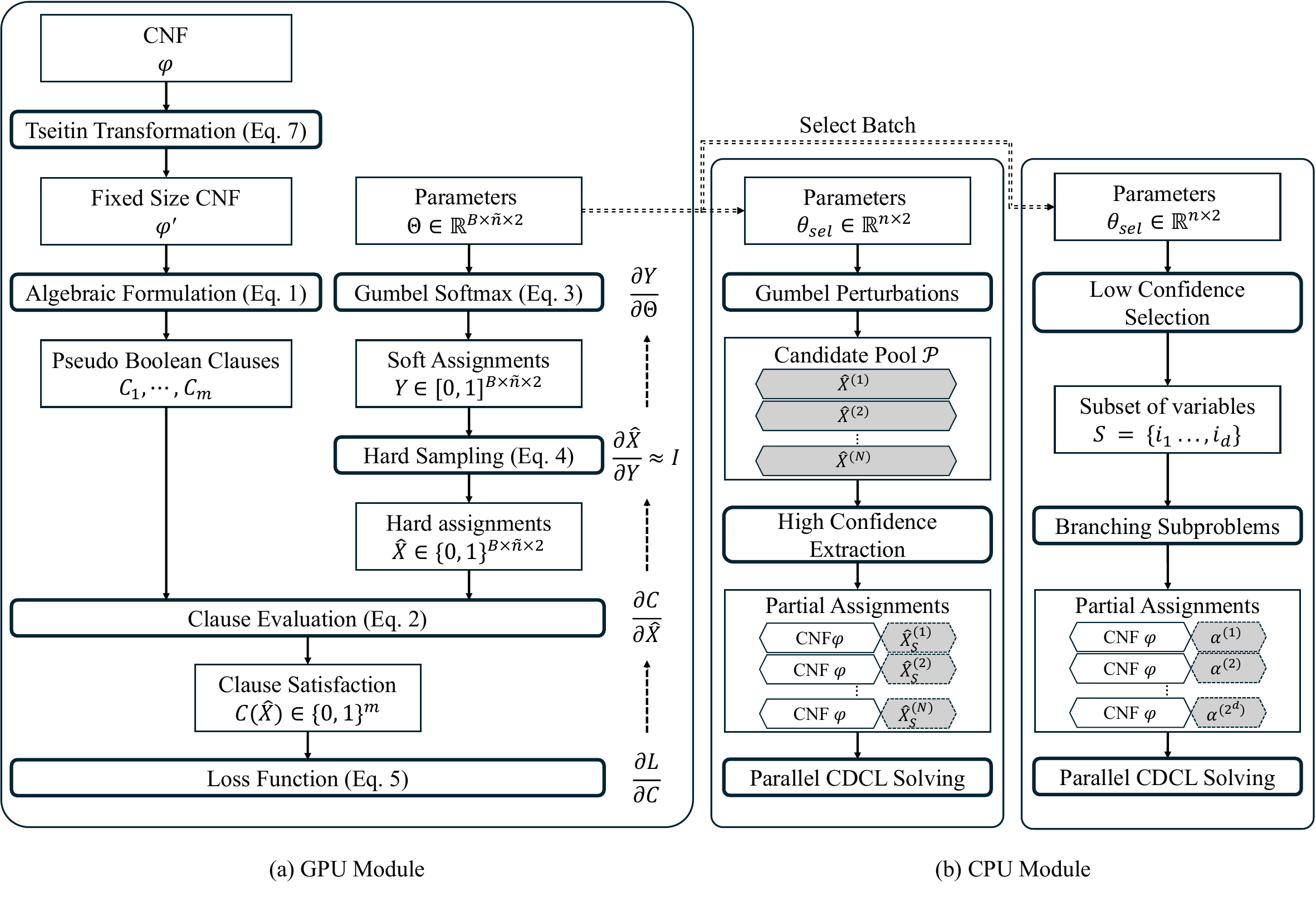}
\caption{
Overview of \textit{GaloisSAT} architecture and solving flow.
(a) \textbf{GPU Module}. Preprocessing and the forward pass for clause evaluation are indicated by solid arrows, while dashed arrows denote backpropagation enabling gradient-based optimization.
(b) \textbf{CPU Module}. Parallel CDCL solving, using high-confidence partial assignments for SAT and branching on low-confidence variables for UNSAT to guarantee completeness.
}
\label{fig:framework}
\end{figure}

Our framework is designed to fully exploit GPU resources, addressing the sequential bottleneck of conventional CPU-based SAT solvers. In this GPU-CPU hybrid architecture, the GPU model predicts variable assignments through gradient-based optimization, while CPUs leverage these assignments to perform parallel CDCL search for both satisfiable and unsatisfiable instances. As shown in Fig.~\ref{fig:framework}a, each CNF instance is first transformed into a fixed-size CNF through Tseitin preprocessing~\cite{tseitin1983complexity}, which unifies clause evaluation across all samples. Clause evaluation is defined in algebraic form, where Boolean operations are reformulated as differentiable polynomial expressions, thus enabling continuous and differentiable evaluation. Each Boolean variable is parameterized by a learnable logit value $\Theta \in \mathbb{R}^{B \times \tilde{n} \times 2}$, where $B$ is the batch size, $\tilde{n}$ is the number of variables after the Tseitin transformation (including auxiliary variables), and the final dimension corresponds to the Boolean states. In the figure, $\hat{X}$ is expressed in a one-hot form for clarity.

Applying continuous relaxation directly to finite-field algebraic representations of Boolean logic fails to preserve the underlying logical dependencies, leading to inaccurate modeling of clause relationships.
Therefore, our model performs a discrete forward pass for accurate clause evaluation while remaining differentiable during backpropagation via the reparameterization trick introduced in Section~\ref{gpu_main}. The objective is to maximize the number of satisfied clauses, allowing the model to learn variable logits accordingly. After training, among all training batches, we select the logits $\theta_{sel}$ from the batch with the minimal clause loss for exploration, and keep the logits for the original $n$ variables. These logits are used to initialize both the SAT and UNSAT frameworks. Unlike data-driven learning approaches that require large labeled datasets, our method is entirely \textbf{data-free}, performing instance-specific optimization based purely on the constraints encoded in the CNF formula.

For \textbf{satisfiable instances}, Fig.~\ref{fig:framework}b, independent Gumbel perturbations are added to $\theta_{sel}$ to sample $N$ full assignments, each associated with a confidence score. Then, each candidate's high-confidence variables are injected as unit clauses into the original CNF, enabling aggressive unit propagation that dramatically reduces the search space, as detailed in Section~\ref{sat}. The resulting $N$ different CNFs are solved in parallel by CDCL threads, terminating the others once a satisfiable solution is found.

Furthermore, our framework also extends to \textbf{unsatisfiable instances}. The confidence scores derived from $\theta_{sel}$ provide a measure of uncertainty for each variable. We select the $d$ least confident variables, interpreting them as the most ambiguous decision points in the problem. These variables serve as branching points for parallel exploration. Inspired by Shannon expansion~\cite{shannon1938symbolic}, we consider all $2^d$ possible assignments to the selected variables and construct $2^d$ subproblems by inserting the corresponding unit clauses. The resulting CNF variants are solved concurrently across $2^d$ CDCL threads in a cube-and-conquer~\cite{heule2011cube} fashion, thus reducing the computational burden of exhaustive UNSAT search, as detailed in Section~\ref{unsat_branching}.

\subsection{Differentiable SAT Formulation}\label{gpu_main}


\paragraph{Differentiable SAT Encoding via Finite Field Algebra}
While CNF provides a convenient representation for Boolean satisfiability, we extend this representation into an finite field algebraic model that enables differentiable reasoning. Each Boolean variable $x, y \in \{0, 1\}$ is treated as a pseudo Boolean variable, and logical operations are reformulated as follows:
\begin{align}\label{eq.algebraic}
\lnot x \;&=\; 1 - x, \nonumber\\
x \wedge y \;&=\; x\cdot y,\\
x \vee y \;&=\; x + y - x\cdot y.\nonumber
\end{align}

 
The algebraic pseudo-Boolean encoding is essential for enabling gradient-based optimization over Boolean structures. Although the variables remain discrete Boolean, representing logical relations through algebraic expressions rather than symbolic logical operators (non-differentiable) makes the formulation differentiable. 
This differentiable formulation enables SAT solving to be reformulated as an optimization problem without relaxing the Boolean domain. 
Moreover, by repeatedly applying De Morgan's laws (e.g., $\lnot(x \vee y) = \lnot x \wedge \lnot y$), any Boolean formula in CNF can be equivalently represented within this algebraic framework. This property ensures that arbitrary clause structures can be handled in a differentiable manner.


Specifically, to illustrate the encoding of SAT problems, we first consider the standard CNF formulation, where each clause represents a disjunction of $u$ literals. We then define each clause modeling as follows:
\begin{equation}\label{eq.eval}
    C = 1 - \prod_{i}^{u}(1 - s_i),\quad \text{where }C \in \{0,1\} \impliedby s_i \in \{0, 1\},
\end{equation}
where each $s_i \in \{0, 1\}$ denotes the Boolean evaluation of the $i$-th literal, i.e., $s_i = 1$ if the literal is satisfied under the current assignment, and $s_i = 0$ otherwise. Thus, $C \in \{0, 1\}$ serves as the satisfiability indicator of the clause, i.e., $C = 1$ when the clause is satisfied, and $C = 0$ otherwise. 



Finally, to enable gradient-based learning, variable assignments are defined in terms of differentiable parameters. Therefore, each Boolean variable $x_i$ is parameterized by a learnable logit vector $\boldsymbol{\theta}_i = [\theta_{i,0}, \theta_{i,1}] \in \mathbb{R}^2$, representing the unnormalized scores for the two Boolean states. In practice, these logits are instantiated in batch form as $\Theta \in \mathbb{R}^{B \times \tilde{n} \times 2}$, where each batch is independently and randomly initialized under our dataless training setup, allowing for unbiased exploration of the variable space.


\paragraph{Gradient-based Optimization with Gumbel-Softmax}

However, purely soft relaxation fails to preserve the logical structure of clauses. Representing each variable as a fractional probability distribution results in clause satisfaction values that do not reflect actual Boolean consistency (e.g., for an OR clause with two variables each assigned 0.5, the relaxed satisfaction evaluates to $1 - 0.5 \cdot 0.5 = 0.75$, overestimating the actual Boolean consistency).

To resolve this, we employ the Gumbel-Softmax reparameterization~\cite{gumbel1954statistical, jang2017categoricalreparameterizationgumbelsoftmax, maddison2017concretedistributioncontinuousrelaxation, liu2024differentiablecombinatorialschedulingscale}, which enables differentiable sampling from a categorical distribution. Specifically, independent Gumbel noise $g_{i,j}$ is added to each logit, followed by a temperature-controlled softmax:
\begin{equation}\label{eq.gumbel}
\mathbf{y}_{i,j} \;=\; \frac{\exp\!\big((\theta_{i,j}+g_{i,j})/\tau\big)}
{\sum_{t\in\{0,1\}} \exp\!\big((\theta_{i,t}+g_{i,t})/\tau\big)},
\quad j\in\{0,1\}, \; g_{i,j}\stackrel{\text{i.i.d.}}\sim \mathrm{Gumbel}(0,1).
\end{equation}

where each $\mathbf{y}_i = [y_{i,0}, y_{i,1}] \in [0,1]^2$ is a differentiable relaxation of the one-hot encoding for the variable $x_i$. As $\tau \to 0$, the distribution sharpens and approaches a categorical one-hot sample, while for larger $\tau$, it remains smoother and supports gradient flow.  

In the forward pass, we use the straight-through (ST) variant, applying a discrete Boolean assignment:
\begin{equation}\label{eq.argmax}
\hat{\mathbf{x}}_i = \argmax_j \mathbf{y}_{i,j},
\quad \hat{\mathbf{x}}_i \in \{0,1\},
\end{equation}
while the backward pass substitutes its gradient with that of the soft relaxation $\mathbf{y}_i$, i.e., $\frac{\partial \mathcal{L}}{\partial \hat{\mathbf{x}}_i}\;\approx\; \frac{\partial \mathcal{L}}{\partial \mathbf{y}_i}$. This approximation allows gradients to pass through the non-differentiable $\argmax$ operation.


In our formulation, we do not directly encode the global conjunction of all clauses (the \textsc{and} operation) as a product, since a single unsatisfied clause would reduce the product to zero and eliminate gradient information. Instead, motivated by the MaxSAT~\cite{hansen1990algorithms} formulation, we adopt a differentiable objective that maximizes the number of satisfied clauses. Here, each $C_t \in \{0,1\}$ indicates whether the $t$-th clause  is satisfied under the sampled Boolean assignment. Let $m$ be the total number of clauses in the CNF formula after applying Tseitin transformation. The loss function is constructed as a sum of differentiable clause evaluations as
\begin{equation}\label{eq.loss}
    \mathcal{L} = - \sum_{t=1}^m C_t,
\end{equation}
so that minimizing $\mathcal{L}$ corresponds to maximizing the number of satisfied clauses. As a result, the model preserves alignment with the discrete MaxSAT semantics while remaining trainable with gradient-based methods.

\paragraph{Clause Normalization for Vectorization}\label{normalization}


Although the clause evaluation in Eq.~\ref{eq.eval} is differentiable, the multiplicative term in Eq.\ref{eq.eval} must have the same length for all clauses to enable vectorized GPU computation. To overcome this limitation, we normalize clause length through the Tseitin transformation~\cite{tseitin1983complexity} before training. The Tseitin transformation introduces auxiliary variables and expands the formula linearly with respect to its original size, while preserving satisfiability.

To transform a long disjunctive clause into a fixed-size CNF form (e.g., 3-CNF), we apply a chain encoding~\cite{biere2009handbook} that preserves satisfiability. Given a clause of length $u$, 
\begin{equation}
    C = (l_1 \vee l_2 \vee \cdots \vee l_u),
\end{equation}
we introduce auxiliary variables $f_1, f_2, \dots, f_{u-3}$ and encode $C$ as a conjunction of 3-literal clauses:
\begin{align}
\psi_C ={}&
(l_1 \vee l_2 \vee f_1) \;\wedge\; \nonumber\\
&(\neg f_1 \vee l_3 \vee f_2) \;\wedge\; \nonumber\\
&(\neg f_2 \vee l_4 \vee f_3) \;\wedge\; \\
&\qquad\qquad\vdots \nonumber\\
&(\neg f_{u-3} \vee l_{u-1} \vee l_u), \nonumber
\end{align}
where $\psi_C$ denotes the set of derived fixed-size clauses corresponding to the original clause $C$. This transformation is equisatisfiable with $C$, and the number of auxiliary variables depends on the chosen fixed clause size. For clauses shorter than the fixed size (e.g., $u<3$), literals are duplicated as needed to match the target clause length.
\begin{equation}
C \text{ is satisfiable } \;\Leftrightarrow\;
\psi_C \text{ is satisfiable.}
\end{equation}
Intuitively, the auxiliary variables $f_i$ propagate the  ``unsatisfied'' state of previous literals through the chain, allowing $\psi_C$ to correctly reflect the satisfaction of the original u-literal clause whenever at least one literal is true.

By applying the above transformation to every clause $C \in \varphi$, we obtain the fixed-size CNF $\varphi'$. Since every clause in $\varphi'$ shares the same algebraic structure and normalized clause length, the model can uniformly evaluate clause satisfiability using the differentiable formulation in Eq.~\ref{eq.eval}. Although the forward pass operates on discrete assignments ($s_i \in \{0,1\}$), differentiability is preserved during backpropagation via the straight-through estimator, ensuring smooth gradient flow through the clause evaluation. With this formulation, all clauses can be tensorized and evaluated in a single vectorized operation. 
This extends naturally across batches, each containing distinct variable assignments, enabling batch-wise clause evaluation on the GPU through parallel linear algebra computation. 
A detailed example of clause normalization is included in Appendix.

\subsection{GaloisSAT Solving Flow}
In our experimental setup, instances were pre-classified into SAT and UNSAT categories to evaluate each framework separately. However, in practical scenarios where the satisfiability status is unknown, both SAT and UNSAT frameworks can be executed concurrently using separate solver threads, maintaining the same overall parallelism.
\subsubsection{Partial Assignment Extraction for SAT Instances}\label{sat}
\paragraph{Sampling via Multiple Gumbel Perturbations}

After training with the Gumbel-Softmax loss, we obtain optimized logits for all variables. For satisfiable instances, the MaxSAT-based optimization yields solutions that are heuristically close to SAT, providing promising candidates for further refinement. As a result, the learned logits often concentrate around regions that are likely to contain feasible SAT solutions. To promote broader exploration and avoid overfitting, we instead select the most uncertain batch, the one with the highest loss value. We construct a candidate pool $\mathcal{P}$, which contains both sampled assignments and their associated confidence scores. Each candidate is generated by sampling from the Gumbel–Softmax distribution $N$ times, where independent Gumbel noise is added to the logit vector of each variable. After adding the noise, we apply the argmax operation (Eq.~\ref{eq.argmax}) to obtain a discrete Boolean assignment for every variable, resulting in a set of $N$ sampled full assignments and their corresponding confidence vectors: 
\begin{equation}\label{eq.pool}
    \mathcal{P} = \{(\hat{\mathbf{x}}^{(1)}, \mathbf{c}^{(1)}), (\hat{\mathbf{x}}^{(2)}, \mathbf{c}^{(2)}), \dots, (\hat{\mathbf{x}}^{(N)}, \mathbf{c}^{(N)})\},
\end{equation}
where each $\hat{\mathbf{x}}^{(k)} = [\hat{x}^{(k)}_1, \ldots, \hat{x}^{(k)}_n] \in \{0,1\}^n$ represents the $k$-th full Boolean assignment, and $c^{(k)} \in [0, 1]^n$ stores the corresponding soft confidence values, reflecting the perturbed logits' probabilities. Here, $n$ refers to the number of variables in the original problem $\varphi$ before the Tseitin transformation, excluding auxiliary variables, so that each $\hat{\mathbf{x}}^{(k)}$ represents a valid assignment in the original variable space. The parameter $N$ is matched to the number of available CPU threads, allowing each candidate assignment can be evaluated in parallel during the solving stage (see Section~\ref{parallel}).

\paragraph{Confidence-guided Partial Assignment Encoding}
For each sampled full assignment $\hat{\mathbf{x}}^{(k)}$, we identify a subset of high-confidence variables to construct a partial assignment. The corresponding confidence vector $\mathbf{c}^{(k)}$ is sorted in descending order, and the top $|S|$ variables are selected. Let $S^{(k)} \subseteq \{1,\dots, N\}$ denote the index set of these selected high-confidence variables for the $k$-th candidate. A fixed selection size $|S|$ is uniformly applied to all $N$ candidates in the pool $\mathcal{P}$, so that each partial assignment $\mathbf{x}_S^{(k)} = [\hat{x}_i^{(k)}]_{i \in S^{(k)}}$ contains the same number of fixed variables.

For each partial assignment $\mathbf{x}_S^{(k)}$, we construct an augmented CNF formula by encoding the selected literals into unit clauses. Each selected variable $x_i \in S^{(k)}$ with a predicted value $\hat{x}_i^{(k)} \in \{0,1\}$ is represented by the corresponding literal $\ell_i^{(k)}$:
\begin{equation}
    (\ell_i^{(k)}) \quad \text{where } 
    \ell_i^{(k)} =
    \begin{cases}
        x_i, & \text{if } \hat{x}^{(k)}_i = 1, \\
        \lnot x_i, & \text{if } \hat{x}^{(k)}_i = 0.
    \end{cases}
\end{equation}
The resulting augmented formula is 
\begin{equation}
    \varphi_\text{aug}^{(k)} = \varphi \;\wedge\; \bigwedge_{i \in S^{(k)}} \ell_i^{(k)}.
\end{equation}
This process produces $N$ augmented CNF instances ${\varphi{\text{aug}}^{(1)}, \dots, \varphi_{\text{aug}}^{(N)}}$, each encoding a distinct high-confidence partial assignment. These instances are distributed across $N$ CPU threads for parallel CDCL solving, where the injected unit clauses trigger early unit propagation and clause simplification, significantly reducing the search space while preserving flexibility in regions of low confidence.

\subsubsection{Guided Branching for UNSAT Instances}\label{unsat_branching}

For unsatisfiable instances, we adopt a branching procedure that partitions the search space into disjoint subproblems. Concretely, given a CNF formula  $\varphi$, we select a subset of variable indices $S = \{i_1, \dots, i_d\}$, corresponding to the variable vector $\mathbf{x}_S = (x_{i_1}, \dots, x_{i_d})$. By Shannon expansion~\cite{shannon1938symbolic} on the selected variables, the original formula is satisfiable if and only if at least one branch is satisfiable. Hence, if all branches are unsatisfiable, the original instance is unsatisfiable (see Lemma~\ref{lemma:shannon}).

\begin{lemma}\label{lemma:shannon}
If $\varphi(x_1, \dots, x_n)$ is a CNF formula and let $S = \{i_1, \dots, i_d\} \subseteq \{1, \dots, n\}$ be any subset of $d$ variables, then
\begin{equation}\label{eq.unsat}
\varphi \;\equiv\; 
\bigvee_{\alpha\in\{0,1\}^d}
\Big( \;\bigwedge_{r=1}^{d} \ell_{i_r}^{(\alpha_r)} \;\wedge\;
\varphi\!\upharpoonright_{\mathbf{x}_S=\alpha} \Big),
\end{equation}
where $\ell_{i_r}^{(1)} := x_{i_r}$, $\ell_{i_r}^{(0)} := \lnot x_{i_r}$, and $\varphi\!\upharpoonright_{\mathbf{x}_S=\alpha}$ is the restriction of $\varphi$ under the assignment $\mathbf{x}_S = \alpha$. If all subformulas $\varphi\!\upharpoonright_{\mathbf{x}_S=\alpha}$ are unsatisfiable for every $\alpha\in\{0,1\}^d$, then $\varphi$ is unsatisfiable.
\end{lemma}

Each branch corresponds to one of the $2^d$ possible Boolean assignments of the selected variables, yielding a restricted subformula of the form $(\bigwedge_{r=1}^{d} \ell_{i_r}^{(\alpha_r)}) \wedge \varphi\!\upharpoonright_{\mathbf{x}_S=\alpha}$, where $\alpha \in \{0,1\}^d$ denotes a particular assignment configuration. In practice, the choice of branching variables critically affects efficiency. Rather than selecting arbitrary variables, we leverage our trained model to identify $d$ variables with the \textit{lowest confidence}, where confidence is derived from the logits' probabilities. Such variables are the most uncertain under the learned distribution. Thus, branching on them is expected to maximize the reduction of ambiguity in the search space. The resulting subinstances are then distributed across available CPU threads for parallel solving, enabling the solver to efficiently explore multiple regions of the search space simultaneously. A more detailed description of the parallel execution strategy is provided in Section~\ref{parallel}.

\subsubsection{Parallel Symbolic Solving on CPUs}\label{parallel}


For satisfiable instances, each augmented CNF formula $\varphi_\text{aug}^{(k)}$ constructed from the candidate pool $\mathcal{P}$ is assigned to a separate CPU thread for CDCL solving. The solving process follows the existential semantics implied by Lemma~\ref{lemma:shannon}: the original problem is satisfied if at least one subinstance $\varphi_\text{aug}^{(k)}$ is satisfiable. Therefore, as soon as any CPU thread returns SAT, all remaining processes are immediately terminated, and the corresponding solution is reported. 
In contrast, for unsatisfiable instances, the process adheres to universal semantics. According to Lemma~\ref{lemma:shannon}, the original instance is unsatisfiable only if all branched subinstances are unsatisfiable. Hence, the system waits for every thread to complete, and termination is declared only after all $2^d$ parallel solvers have returned UNSAT. 
This hybrid GPU-CPU coordination fully exploits available parallelism. The GPU provides confidence-guided decomposition of the search space, while the CPU ensemble executes exact symbolic solving in parallel, terminating early when satisfiability is established or exhaustively verifying unsatisfiability.

%% file: 4-Experiments.tex
\section{Experiments}\label{experiments}

This section presents the main results of \textit{GaloisSAT} evaluated with SAT Competition 2024 benchmarks~\cite{heule2024proceedings}. The details of our system configurations and algorithmic hyperparameters are included in Appendix. For the reported experiments, instances were pre-classified into SAT and UNSAT categories to separately evaluate solver performance.

\subsection{SAT Performance against SOTA Solvers}

\begin{figure}[ht]
    \centering
    \begin{subfigure}[t]{0.48\textwidth}
        \centering
        \includegraphics[width=\linewidth]{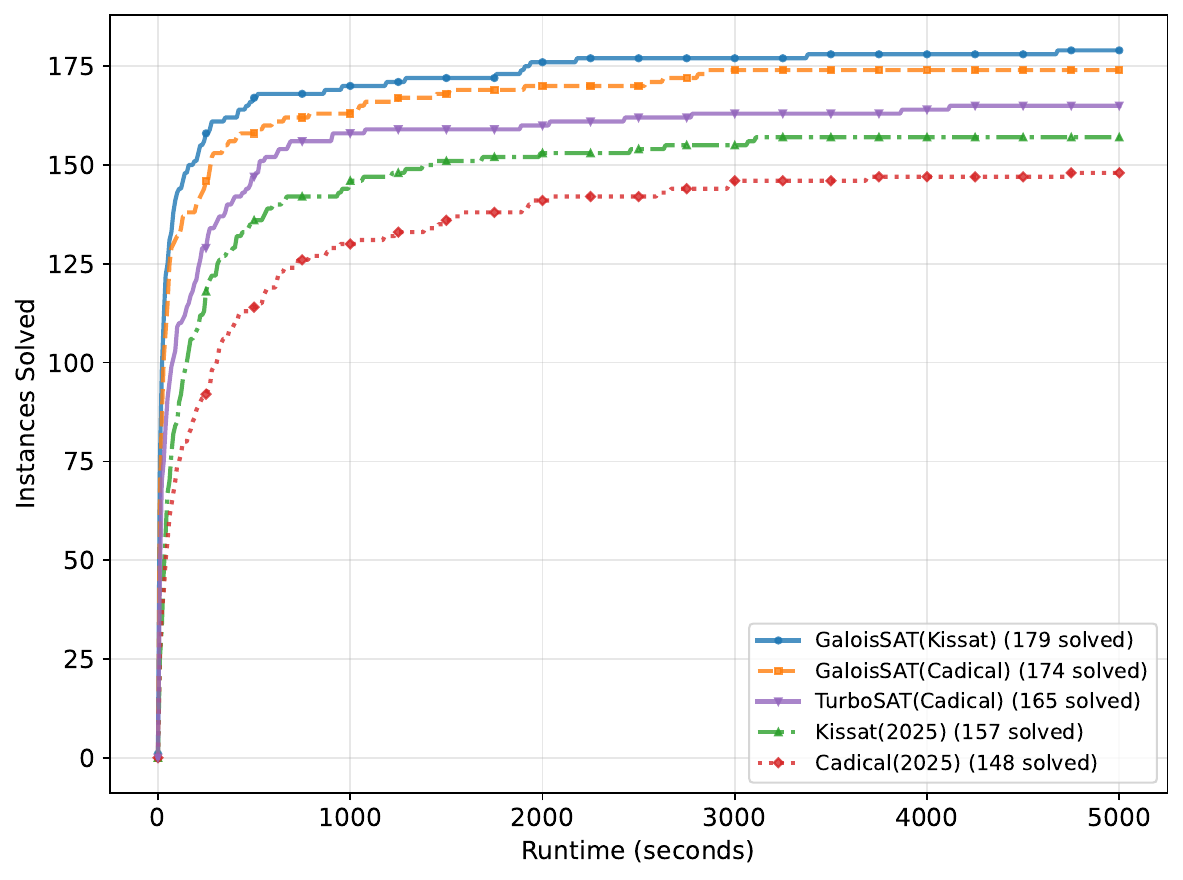}
        \caption{Cumulative solved instances over time}
        \label{fig:cumulative_sat}
    \end{subfigure}
    \hfill
    \begin{subfigure}[t]{0.48\textwidth}
        \centering
        \includegraphics[width=\linewidth]{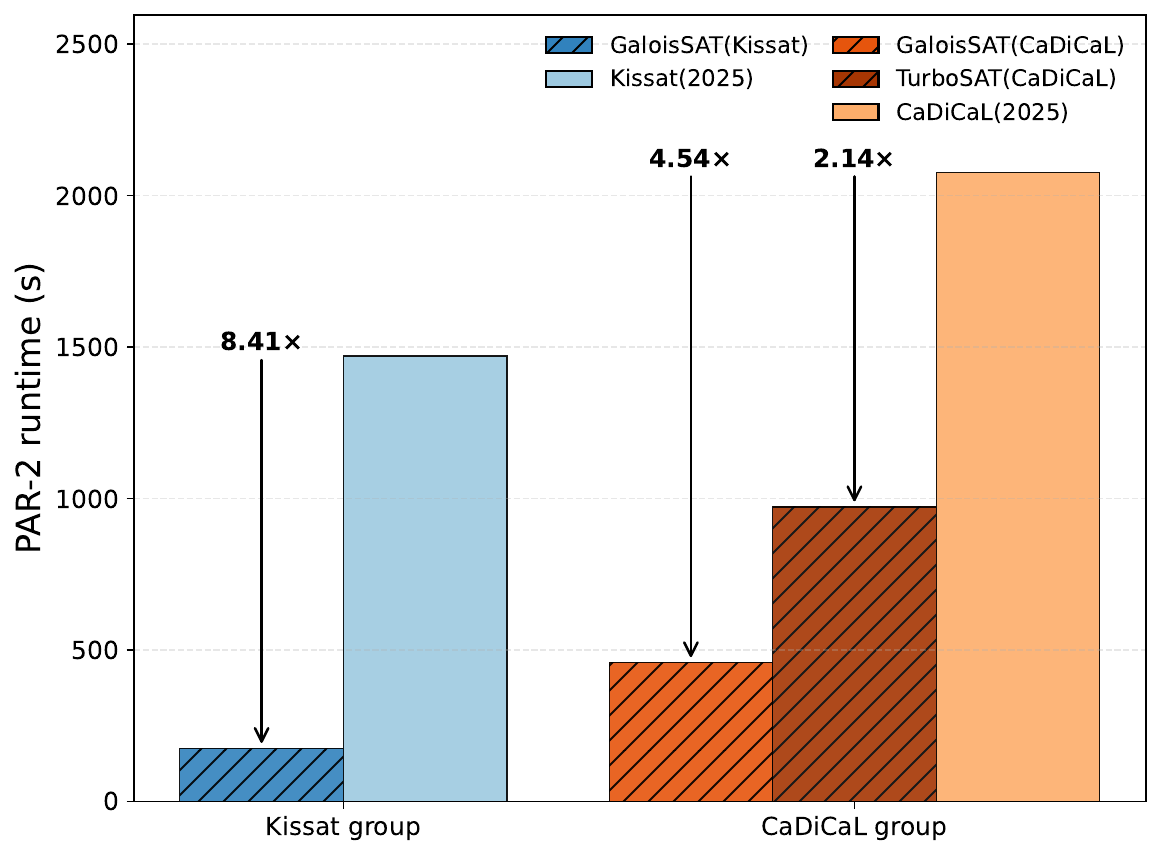}
        \caption{PAR-2 score improvement}
        \label{fig:par2_sat}
    \end{subfigure}
    \caption{Performance comparison between baseline solvers {\textit{Kissat}, \textit{CaDiCaL}} and their {\textit{Galois}}-augmented variants, as well as \textit{TurboSAT(CaDiCaL)}, on the 179 satisfiable instances from the SAT Competition 2024 dataset. *Note that the PAR-2 scores for \textit{GaloisSAT} account for the total end-to-end runtime of the entire GPU+CPU solving flow.}
    \label{fig:result_sat}
\end{figure}

We first compare our proposed method with single-threaded, state-of-the-art CDCL solvers on SAT instances from the SAT Competition 2024 benchmark. For this experiment, we parallelized the search using 100 CPU threads ($N=100$ in Eq.~\ref{eq.pool}). Figure~\ref{fig:cumulative_sat} shows the cumulative number of solved instances over time. The curves clearly demonstrate that our hybrid solvers, \textit{GaloisSAT(Kissat)} and \textit{GaloisSAT(CaDiCaL)}, consistently outperform their standalone counterparts, as well as \textit{TurboSAT(CaDiCaL)}, across all time ranges. In particular, within the first 200 seconds, our approach solves substantially more instances, highlighting its efficiency in reaching satisfiable assignments rapidly. On the 179 \textit{satisfiable} instances, at the end of the 5,000-second timeout, \textit{GaloisSAT(Kissat)} and \textit{GaloisSAT(CaDiCaL)} solved 179 and 174 cases, respectively, whereas the original \textit{Kissat} and \textit{CaDiCaL} solved only 157 and 148. This result indicates that our method not only accelerates solving but also improves robustness, solving a greater number of SAT instances within the same overall runtime. 

Next, to quantitatively and jointly evaluate both efficiency and robustness, we adopted the PAR-2 metric as our primary performance measure. This metric penalizes timeouts while rewarding faster solutions, providing a balanced assessment of overall solver quality. The results, depicted in Figure~\ref{fig:par2_sat}, show that lower PAR-2 values correspond to superior performance. Our \textit{Galois}-enhanced variants consistently surpassed their original backbones, achieving up to 8.41$\times$ speedup over \textit{Kissat} and 4.54$\times$ speedup over \textit{CaDiCaL}. Notably, while \textit{TurboSAT(CaDiCaL)} improved upon \textit{CaDiCaL} by 2.14$\times$, \textit{GaloisSAT(CaDiCaL)} delivers even greater gains, demonstrating the effectiveness of our approach. The precise PAR-2 scores corresponding to these results are provided in Table~\ref{tab:comparison}.
Our evaluation shows that \textit{GaloisSAT} provides a scalable and efficient plug-in framework that is compatible with modern SAT solvers and extensible to future solver architectures.

\begin{figure}[t]
    \centering
    \begin{subfigure}[t]{0.48\textwidth}
        \centering
        \includegraphics[width=\linewidth]{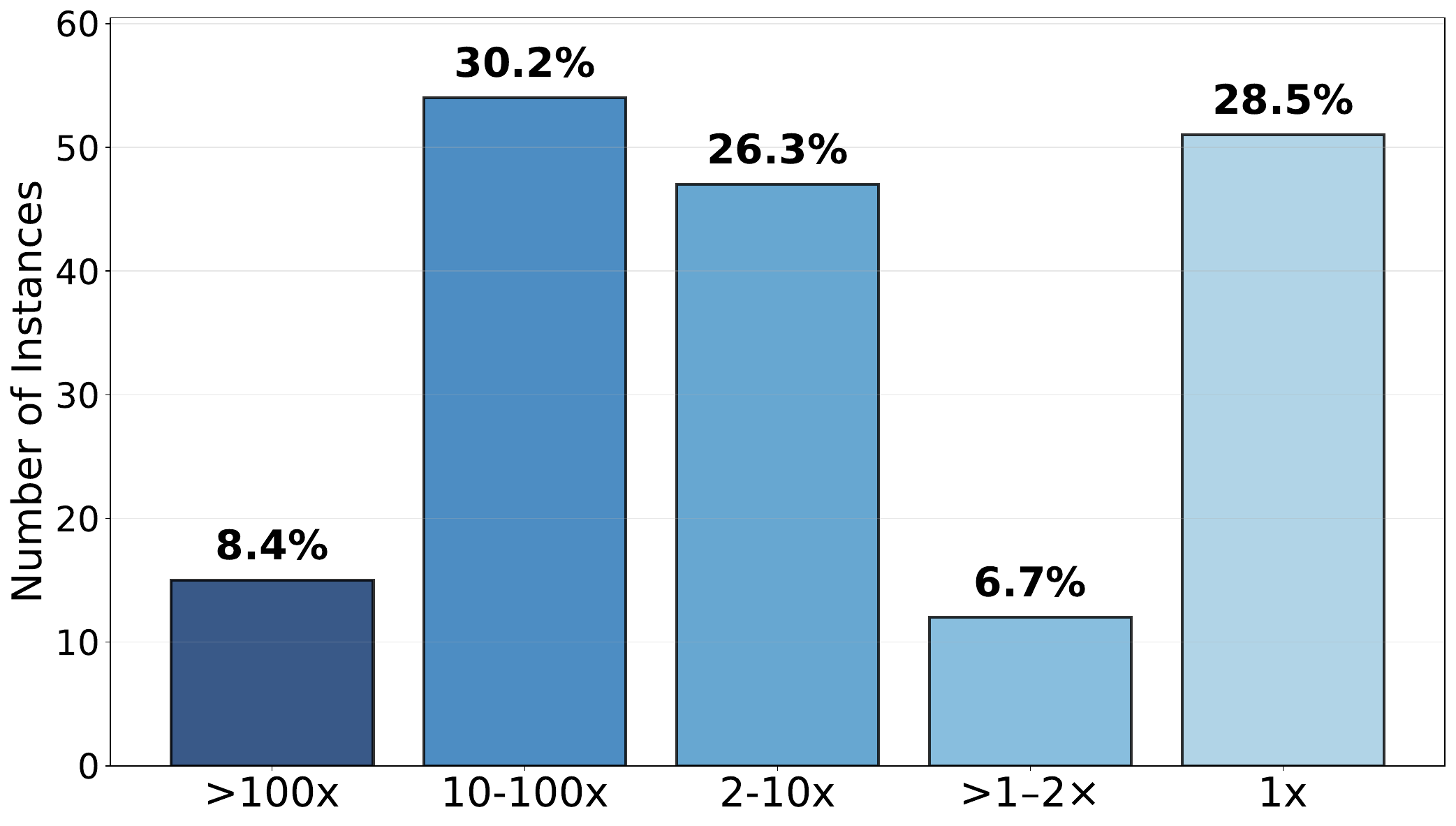}
        \caption{\textit{GaloisSAT(Kissat)}}
        \label{fig:speedup_kissat}
    \end{subfigure}
    \hfill
    \begin{subfigure}[t]{0.48\textwidth}
        \centering
        \includegraphics[width=\linewidth]{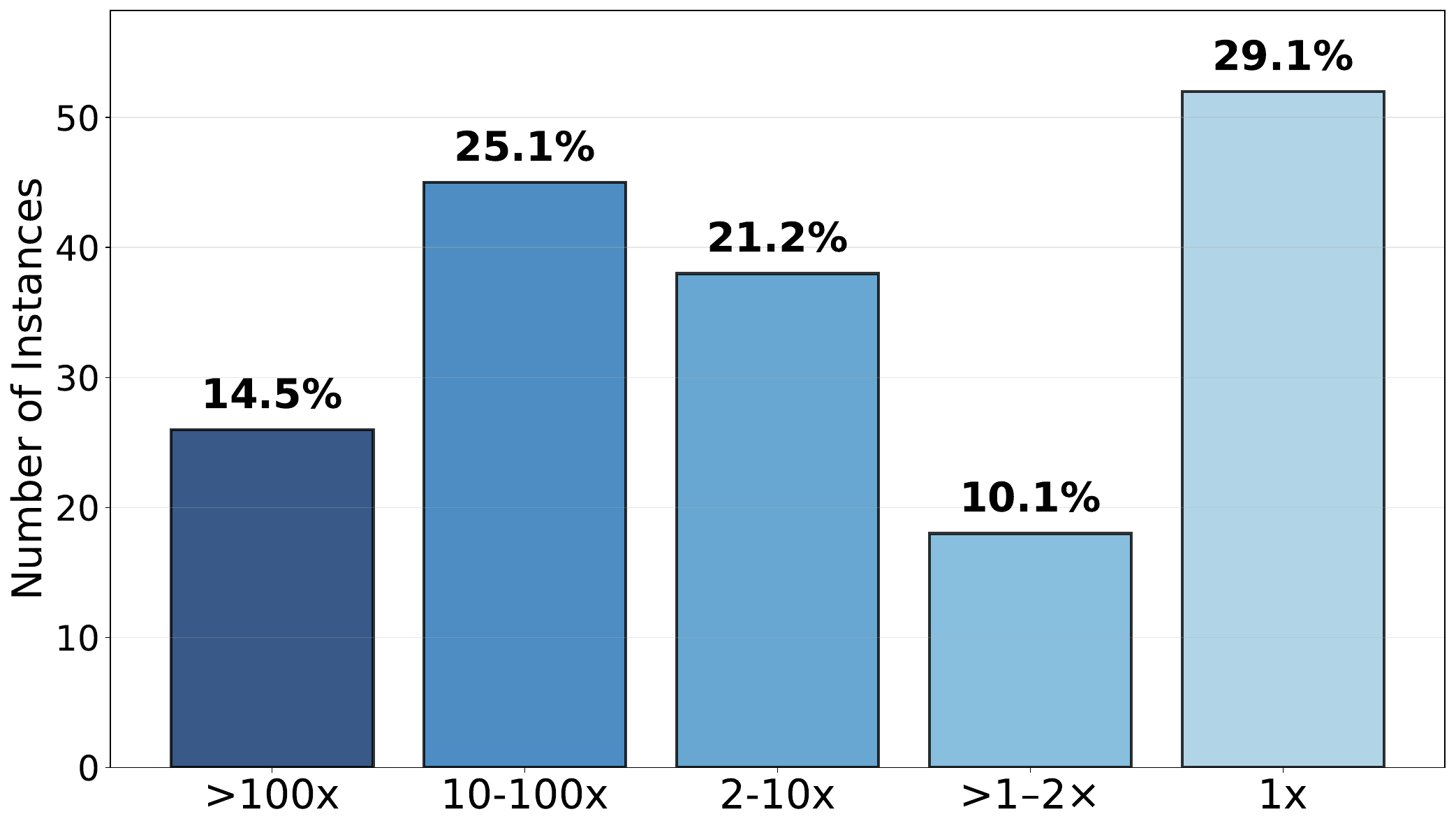}
        \caption{\textit{GaloisSAT(CaDiCaL)}}
        \label{fig:speedup_cadical}
    \end{subfigure}
    \caption{Distribution of speedups on satisfiable instances from the SAT Competition 2024 benchmarks. Both (a)~\textit{GaloisSAT(Kissat)} and (b)~\textit{GaloisSAT(CaDiCaL)} achieve substantial improvements on over 70\% of the instances.
    }
    \label{fig:speedup}
\end{figure}

We next analyze how much speedup it achieves across all instances. Figure~\ref{fig:speedup} illustrates the speedup distributions of our \textit{Galois}-augmented solvers relative to their original backbones. For \textit{Galois(Kissat)}, 8.4\% of the satisfiable benchmarks achieved over 100$\times$ speedup, and 30.2\% fell within the 10-100$\times$ range, while for \textit{Galois(CaDiCaL)}, 14.5\% and 25.1\% of the instances achieved those respective levels. In total, more than 70\% of all satisfiable instances exhibit noticeable acceleration. Many of the extreme speedup values arise from instances that timed out in the baseline solvers, which are counted as 10,000 seconds under the PAR-2 definition. The cases showing around $1\times$ speedup typically correspond to easy or trivially satisfiable problems, where the GPU overhead dominates, or to large but simple instances that require little search. Such cases are not problematic since we can include one original CNF (without any injected unit clauses) among the CPU threads, ensuring that the baseline CDCL solver performance serves as a guaranteed lower bound. In contrast, on challenging instances, our approach delivers speedup of several orders of magnitude, marking a clear breakthrough over existing solvers.

\begin{table}[ht]
\centering
\caption{Performance comparison with SOTA SAT solvers.}
\label{tab:comparison}
\begin{tabular}{l|c|c|c}
\hline
\multicolumn{1}{c|}{\textbf{Solver}} &
\multicolumn{1}{c|}{\textbf{Avg. PAR-2 $\downarrow$}} &
\multicolumn{1}{c|}{\textbf{Solved}} &
\multicolumn{1}{c}{\textbf{Solved Rate}} \\
\hline
\textit{\textbf{GaloisSAT(Kissat)}} & \textbf{174.79}  & \textbf{179}/179 & 100.0\% \\
\textit{Kissat}                 & 1470.60 & 157/179 & 87.7\%   \\
\hline
\textit{\textbf{GaloisSAT(CaDiCaL)}} & \textbf{457.76} & \textbf{174}/179 & 97.2\%  \\
\textit{TurboSAT(CaDiCaL)}  & 972.28 & 165/179 & 92.2\% \\
\textit{CaDiCaL}                & 2076.91 & 148/179 & 82.7\% \\
\hline
\end{tabular}
\end{table}


We further analyzed the per-instance behavior of our \textit{Galois}-augmented solvers to better understand how the improvements manifest across different problem difficulties. As detailed in Table~\ref{tab:top-improved}, our approach effectively accelerates even the hard SAT benchmarks, many of which originally required hundreds to thousands of seconds, or even reached the timeout limit to solve. In particular, \textit{GaloisSAT(Kissat)} achieved up to a 6050.34$\times$ speedup, while \textit{GaloisSAT(CaDiCaL)} reached as high as 820.29$\times$, outperforming \textit{TurboSAT(CaDiCaL)} in most cases. For example, on \textit{mdp-32-14-sat}, \textit{TurboSAT(CaDiCaL)} ran for 905.65~s, whereas \textit{GaloisSAT(CaDiCaL)} completed in 12.19~s. 

\begin{table}[t]
\centering
\footnotesize
\setlength{\tabcolsep}{1.5pt}
\begin{subtable}[t]{\linewidth}
\centering
\caption{\textit{GaloisSAT(Kissat)}}\label{tab:top-kissat}
\begin{tabularx}{\linewidth}{L|C{2cm}|C{2.75cm}|C{2.75cm}|C{1.6cm}}
\hline
\multicolumn{1}{c|}{\textbf{Instance}\rule[-1em]{0pt}{2.5em}} &
\multicolumn{1}{c|}{
  \begin{tabular}[c]{@{}c@{}}
    \textbf{\#Variables/} \\
    \textbf{\#Clauses}
  \end{tabular}
}&
\multicolumn{1}{c|}{\textbf{Kissat}} &
\multicolumn{1}{c|}{
  \begin{tabular}[c]{@{}c@{}}
    \textbf{GaloisSAT} \\
    \textbf{(Kissat)}
  \end{tabular}
}&
\multicolumn{1}{c}{
  \begin{tabular}[c]{@{}c@{}}
    \textbf{PAR-2} \\
    \textbf{Speedup}
  \end{tabular}
}\\
\hline
mp1-Nb7T42                      &52731/208936& 5,000(TO) & 1.65 & 6,050.34$\times$ \\
pcmax-scheduling-m15            &2351/13561 & 5,000(TO) & 3.77 & 2,649.36$\times$ \\
pcmax-scheduling-m37         &28830/324346 & 5,000(TO) & 11.04 & 905.47$\times$ \\
x9-12098.sat                 &600/5395 & 555.17 & 0.89 & 622.25$\times$ \\
mdp-28-14-sat                 &792/4386 & 383.94 & 0.88 & 438.44$\times$ \\
rbsat-v760c43649gyes3         &760/43649& 1,378.80 & 3.26 & 422.62$\times$ \\
rbsat-v760c43649gyes7        &760/43649 & 1,683.77 & 4.82 & 349.22$\times$ \\
mdp-32-14-sat                 &1027/5750& 5,000(TO) & 47.20 & 211.86$\times$ \\
pcmax-scheduling-m24 & 24101/255206& 5,000(TO) & 55.40 & 180.50$\times$ \\
ex065\_25                     &74804/393322& 3109.65 & 20.22 & 153.78$\times$ \\
\hline
\end{tabularx}
\end{subtable}
\medskip
\begin{subtable}[t]{\linewidth}
\centering
\caption{\textit{GaloisSAT(CaDiCaL)}}\label{tab:top-cadical}
\begin{tabularx}{\linewidth}{L|C{2cm}|C{1.8cm}|C{1.8cm}|C{1.8cm}|C{1.6cm}}
\hline
\multicolumn{1}{c|}{\textbf{Instance}\rule[-1em]{0pt}{2.5em}} &
\multicolumn{1}{c|}{
  \begin{tabular}[c]{@{}c@{}}
    \textbf{\#Variables/} \\
    \textbf{\#Clauses}
  \end{tabular}
}&
\multicolumn{1}{c|}{\textbf{CaDiCaL}} &
\multicolumn{1}{c|}{
  \begin{tabular}[c]{@{}c@{}}
    \textbf{TurboSAT} \\
    \textbf{(CaDiCaL)}
  \end{tabular}
}&
\multicolumn{1}{c|}{
  \begin{tabular}[c]{@{}c@{}}
    \textbf{GaloisSAT} \\
    \textbf{(CaDiCaL)}
  \end{tabular}
}&
\multicolumn{1}{c}{
  \begin{tabular}[c]{@{}c@{}}
    \textbf{PAR-2} \\
    \textbf{Speedup}
  \end{tabular}
}\\
\hline
mdp-32-14-sat                     &1027/5750 & 5,000(TO) & 905.65 & 12.19 & 820.29$\times$ \\
rbsat-v760c43649gyes3             &760/43649& 2,963.52 & 11.97 & 4.34 & 682.45$\times$ \\
rbsat-v760c43649gyes7             &760/43649& 872.59 & 11.21 &  1.34 & 650.46$\times$ \\
battleship-14-26-sat              &364/2562& 309.26 & 4.65 &  0.58 & 531.01$\times$ \\
002                               &4128/126564& 5,000(TO) & 471.06 & 20.39 & 490.38$\times$ \\
x9-12092.sat                      &600/5399& 273.97 & 96.47 & 0.57 & 477.22$\times$ \\
sgen1-sat-180-100                 &180/432& 5,000(TO) & 12.28 & 21.91 & 456.43$\times$ \\
x9-11077.sat                      &550/4950& 948.47 & 13.49 & 2.11 & 449.11$\times$ \\
x9-10038.sat                      &500/4499& 175.67 & 28.50 & 0.40 & 437.75$\times$ \\
x9-10014.sat                      &500/4506& 182.55 & 4.88 & 0.52 & 349.65$\times$ \\
\hline
\end{tabularx}
\end{subtable}

\caption{Top satisfiable instances showing the largest runtime reductions achieved by \textit{GaloisSAT}. Both \textit{GaloisSAT(Kissat)} and \textit{GaloisSAT(CaDiCaL)} deliver substantial gains, frequently exceeding two-digit speedups and reaching up to three orders of magnitude compared to their backbone solvers. In most cases, \textit{GaloisSAT(CaDiCaL)} also surpasses \textit{TurboSAT(CaDiCaL)}. Entries marked as TO indicate timeouts.}\label{tab:top-improved}
\end{table}

\subsection{UNSAT Performance Improvement}

\begin{figure}[ht]
    \centering
    \begin{subfigure}[t]{0.48\textwidth}
        \centering
        \includegraphics[width=\linewidth]{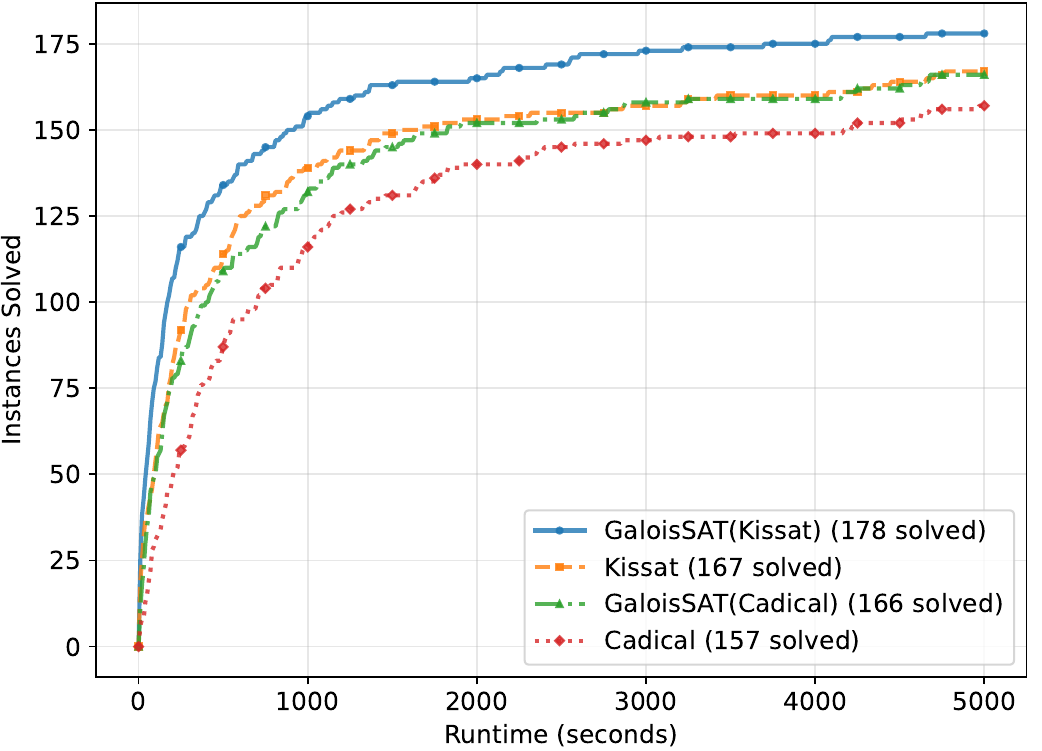}
        \caption{Cumulative solved instances over time}
        \label{fig:cumulative_unsat}
    \end{subfigure}
    \hfill
    \begin{subfigure}[t]{0.48\textwidth}
        \centering
        \includegraphics[width=\linewidth]{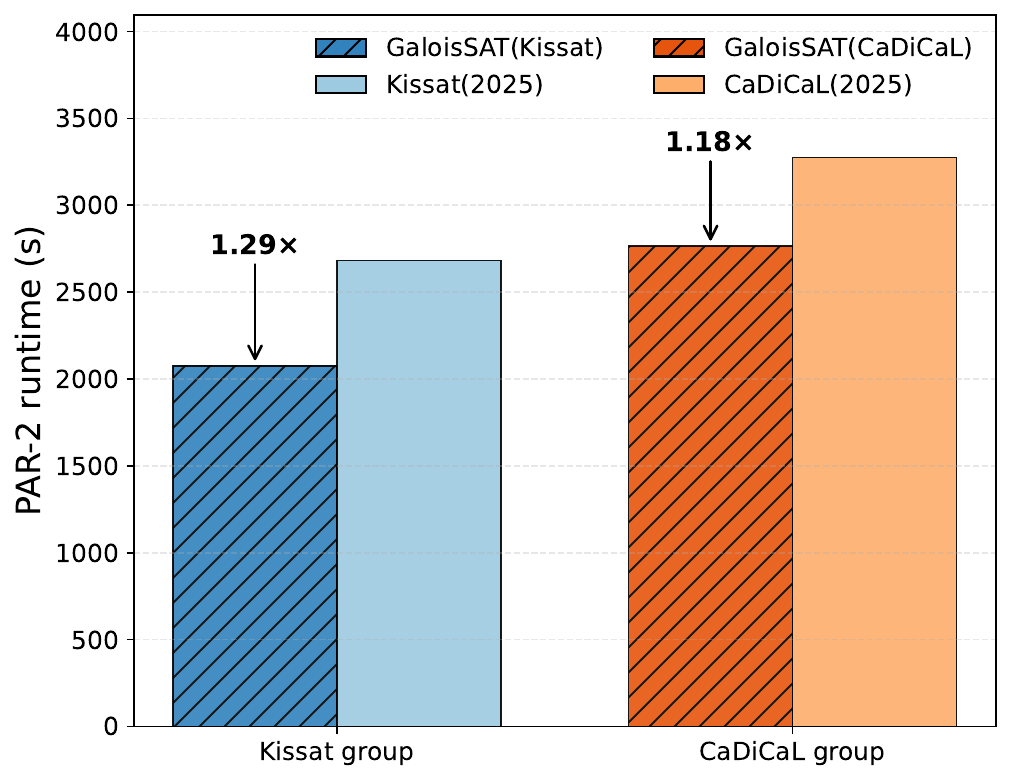}
        \caption{PAR-2 score improvement}
        \label{fig:par2_unsat}
    \end{subfigure}
    \caption{Performance comparison between baseline solvers {\textit{Kissat}, \textit{CaDiCaL}} and their {\textit{Galois}}-augmented variants on the 214 unsatisfiable instances from the SAT Competition 2024 dataset. *Note that the PAR-2 scores for \textit{GaloisSAT} account for the total end-to-end runtime of the GPU+CPU solving flow.}
    \label{fig:result_unsat}
\end{figure}

We evaluated the proposed framework on unsatisfiable instances to assess its effectiveness in accelerating proof-based search. Unlike the satisfiable case, the goal here is to exhaustively verify the absence of any satisfying assignment, which typically requires more computationally intensive clause learning. For this experiment, we parallelized the search using 128 CPU threads ($d=7$ in Lemma~\ref{lemma:shannon}). Figure~\ref{fig:cumulative_unsat} illustrates the cumulative number of solved instances over time on the UNSAT subset of the SAT Competition 2024 benchmarks. Among the 214 unsatisfiable instances, \textit{GaloisSAT(Kissat)} solved 178 cases compared to 167 by \textit{Kissat}, while \textit{GaloisSAT(CaDiCaL)} solved 166 cases compared to 157 by \textit{CaDiCaL} within a 5,000-seconds limit. The cumulative curves show that both Galois-enhanced solvers consistently outperform their backbone solvers throughout the entire solving time, confirming the robustness of our framework in providing UNSAT proofs. Note that, unlike \textit{GaloisSAT}, current SOTA solver \textit{TurboSAT} does not provide UNSAT certification.

In terms of overall efficiency, Figure~\ref{fig:par2_unsat} shows that \textit{GaloisSAT(Kissat)} achieved a PAR-2 score reduction from 2683.55 to 2074.86, corresponding to a 1.29$\times$ improvement. A similar trend holds for \textit{GaloisSAT(CaDiCaL)}, which reduced its PAR-2 score from 3276.10 to 2764.79, achieving a 1.18$\times$ improvement. This demonstrates the general applicability of our framework across both satisfiable and unsatisfiable domains. Notably, when branching variables were chosen arbitrarily instead of by the learned policy, the solvers exhibited more frequent timeouts and overall worse performance than when no branching was applied.

\subsection{Ablation Studies and Analysis}

\subsubsection{GPU Runtime and Hardware Scaling}


The overall runtime of our framework consists of two parts: the GPU-side differentiable learning stage and the CPU-side CDCL solving stage. The latter, which deterministically evaluates candidate assignments generated by the GPU, is relatively fixed and offers limited room for further acceleration. In contrast, the GPU stage dominates the total runtime and presents the greatest opportunity for optimization. The GPU execution time of our method primarily depends on three factors: (i) the size of the SAT instances, (ii) available GPU memory, and (iii) the batch size. In our experiments, the instance size is fixed, and we therefore conduct ablation studies to examine the impact of GPU memory and batch size.


\begin{table}[t]
\centering
\caption{Average GPU runtime on the SAT instances from the SAT Competition 2024 benchmarks across different hardware configurations and training settings.}
\label{tab:gpu_configuration}
\begin{tabular}{l|c|c|c|c}
\hline
\textbf{\# GPUs} & \textbf{Batch} & \textbf{Avg. Load (s)}& \textbf{Avg. Train (s)} &\textbf{Avg. Total GPU (s)} \\
\hline
\multirow{3}{*}{2×A100} 
    & 500  & 1.39& 9.03 &  10.42 \\
    & 1,000  & 1.29& 17.90 & 19.19  \\
    & 3,000  & 1.46& 52.31 & 53.77  \\
\hline
8×A100 & 3,000 &3.45 & 15.24 & 18.69 \\
\hline
\end{tabular}
\end{table}

For extremely large instances, the batch is divided into sub-batches for gradient accumulation due to memory limits, which introduces additional overhead. However, with sufficiently large-memory GPUs, this overhead can be significantly reduced. As the hardware scales from 2$\times$A100 to 8$\times$A100 while fixing batch size as 3,000, the average GPU runtime decreases from 53.77~s to 18.69~s (2.87$\times$), and pure training time from 52.31~s to 15.24~s (3.43$\times$) as shown in Table~\ref{tab:gpu_configuration}. Even for the largest instance (48,505,464 literals, 130,975,382 clauses), the training time decreases substantially, from 884.79~s to 238.18~s, obtaining 3.71$\times$ speedup. This near-linear speedup demonstrates that our framework directly benefits from both higher GPU parallelism and larger memory capacity.

 
To investigate the effect of batch size, we compared three training configurations: a batch size of 500, 1,000, and 3,000, with a fixed number of epochs of 10. 
The smaller batch size improved GPU-side efficiency, reducing GPU execution time by 5.16$\times$ and 2.80$\times$ when using batch sizes of 500 and 1,000, respectively, compared to a batch size of 3,000 (Table~\ref{tab:gpu_configuration}).

We further compared those configurations regarding CPU solving performance across each CDCL solver, as summarized in Table~\ref{tab:batch_ablation}. The overall end-to-end performance consistently exceeded that of the backbone solvers, demonstrating the robustness of our framework across different training settings. For \textit{GaloisSAT(Kissat)}, increasing the batch size improved both the PAR-2 score and the solved rate. \textit{GaloisSAT(CaDiCaL)} failed to solve two instances at a batch size of 3,000, resulting in a higher PAR-2 score. When accounting for the penalty from these unsolved instances and recalculating the PAR-2 for the solved problems ($457.76 - 2 \times 10{,}000 / 179 = 346.03$), the effective PAR-2 is actually lower than that obtained with smaller batch sizes, demonstrating improved performance on the solved instances. This behavior can be attributed to the increased probability that the selected batch corresponds to an outlier as the batch size grows. While \textit{Kissat} benefits from more robust heuristics, \textit{CaDiCaL} is more sensitive to batch size due to suboptimal initializations that can lead to timeouts.

\begin{table}[ht]
\centering
\caption{End-to-end solver performance across different batch sizes, along with baseline results. All \textit{Galois}-enhanced solvers were trained using 2$\times$A100 GPUs with a number of epochs of 10.}
\label{tab:batch_ablation}
\begin{tabular}{l|c|c|c|c}
\hline
\multicolumn{1}{c|}{\textbf{Solver}} &
\multicolumn{1}{c|}{\textbf{Batch}} &
\multicolumn{1}{c|}{\textbf{Avg. PAR-2 $\downarrow$}} &
\multicolumn{1}{c|}{\textbf{Solved}} &
\multicolumn{1}{c}{\textbf{Solved Rate}} \\
\hline
\multirow{3}{*}{\textit{GaloisSAT(Kissat)}} 
    & 3,000 & 174.79  & 179/179 & 100.0\% \\
    & 1,000 & 229.77 & 178/179 & 99.4\% \\
    & 500  & 241.68  & 178/179 & 99.4\% \\
\hline
\multirow{3}{*}{\textit{GaloisSAT(CaDiCaL)}} 
    & 3,000 & 457.76 & 174/179 & 97.2\%  \\
    & 1,000 & 367.50 & 176/179 &  98.3\%  \\
    & 500 & 394.98 & 175/179 & 97.8\%  \\
\hline
\end{tabular}
\end{table}




\subsubsection{Quality Assessment of Partial Assignments}

To further assess the effectiveness of the generated candidate pool, we analyzed the quality of 100 candidates per satisfiable instance from the SAT Competition 2024 benchmark set (a total of 17,900 candidates). In Table~\ref{tab:candidate_analysis}, each candidate was evaluated using \textit{Kissat}. Among them, 78.58\% were satisfiable, 15.44\% reached the timeout limit, 5.59\% were unsatisfiable, and 0.34\% were labeled as unknown. These results confirm that a large majority of the generated candidates correspond to valid SAT solutions, indicating the reliability of our learned initialization. 

\begin{table}[ht]
\centering
\caption{Candidate pool quality for 100 generated candidates per satisfiable instance (179 instances) from the SAT Competition 2024 dataset, evaluated using \textit{Kissat}. Each result type shows the proportion of candidates under this configuration.}
\label{tab:candidate_analysis}
\begin{tabular}{l|c}
\hline
\multicolumn{1}{c|}{\textbf{Result Type}} &
\multicolumn{1}{c}{\textbf{Proportion (\%)}} \\
\hline
Satisfiable (succeeded partial assignments)  & 78.58 \\
Timeout        & 15.44 \\
Unsatisfiable (failed partial assignments)   & 5.59  \\
Unknown     & 0.34  \\
\hline
\end{tabular}
\end{table}

%% file: 2-Preliminary.tex
\section{Related Work}\label{preliminary}




\subsection{Boolean Satisfiability Problem (SAT)}
The \textbf{Boolean Satisfiability Problem (SAT)} is a fundamental problem in computer science and mathematical logic. It asks whether there exists an assignment of truth values to Boolean variables that makes a given logical formula evaluate to true. 
A SAT problem is typically expressed in Conjunctive Normal Form (CNF), where a formula is a conjunction of clauses, and each clause is a disjunction of literals. Consider the CNF formula:
\[
\varphi = (x_1 \lor \lnot x_2) \land (\lnot x_1 \lor x_3).
\]
To satisfy $F$, we must assign truth values to $x_1$, $x_2$, and $x_3$ such that all clauses evaluate to true. One satisfying assignment is:
\[
x_1 = \text{True}, \quad x_2 = \text{False}, \quad x_3 = \text{True},
\]
since both $(x_1 \lor \lnot x_2)$ and $(\lnot x_1 \lor x_3)$ evaluate to true. Determining such assignments efficiently for large formulas lies at the heart of SAT solving.

\subsection{Supervised Learning for SAT Solving}
Supervised learning for SAT solving typically follows two paths: end-to-end prediction or heuristic enhancement. NeuroSAT~\cite{Selsam2019NeuroSAT} introduces an end-to-end message-passing neural network that predicts satisfiability by encoding CNF formulas as bipartite graphs, effectively framing the problem as a binary classification task. 
Building on this paradigm, SATFormer~\cite{shi2023satformer} employs a hierarchical Transformer architecture to capture clause-level dependencies, enabling both SAT and UNSAT prediction, as well as UNSAT core identification. 

In contrast, supervised learning–based heuristic enhancement methods aim to improve specific components of traditional solvers. NeuroCore~\cite{Selsam2019NeuroCore} guides CDCL branching by predicting variables likely to appear in UNSAT cores. 
Neuroback~\cite{Wang2024NeuroBack} predicts promising variable assignments based on patterns observed in satisfying solutions and incorporates them into decision heuristics. Despite these improvements, both categories of supervised methods fundamentally rely on large labeled datasets, which can hinder scalability and limit generalization to unseen problem distributions.

\subsection{Unsupervised Learning for SAT Solving}

Unsupervised learning approaches reformulate SAT solving as a continuous optimization problem. FourierSAT~\cite{kyrillidis2020fouriersat} converts Boolean constraints into multilinear polynomials via Walsh-Fourier expansions, guiding the search through gradient descent.
FastFourierSAT~\cite{Cen2025FastFourierSAT} employs GPU-parallelized Fast Fourier Transform to accelerate the computationally intensive evaluation of elementary symmetric polynomials. However, FastFourierSAT relies on high-precision arithmetic to ensure numerical stability, preventing it from fully exploiting modern GPU features. Distinct from polynomial-based methods, DG-DAGRNN~\cite{amizadeh2018learning} adopts a neural continuous optimization framework. It maps the SAT formula to a Directed Acyclic Graph (DAG) and uses Recurrent Neural Networks (RNNs) to iteratively update continuous variable assignments by minimizing a differentiable unsupervised loss function.


In addition, reinforcement learning has been applied to SAT solving to improve branching heuristics by learning from solver interactions. Graph-Q-SAT~\cite{kurin2020can} uses GNNs to approximate the Q function, guiding variable selection in a CDCL solver. Similarly, Glue Variable Elimination (GVE)~\cite{zhang2021elimination} leverages learned structural insights to identify key variables before applying a deterministic solver with reinforcement learning. The primary bottleneck of these methods is the high inference latency of neural networks compared to lightweight heuristics like VSIDS, which often negates the reduction in search steps.

\subsection{Parallel and Distributed SAT Solving}
To overcome the stagnation of single-core CPU performance, recent research has explored various parallel architectures. ParaFrost~\cite{osama2024parafrost} adopts a hybrid architecture that offloads computationally intensive inprocessing tasks to the GPU, specifically variable elimination and subsumption. By utilizing the GPU's high throughput to simplify the formula structure, it effectively reduces the search space before handing the instance back to a CDCL solver. However, for hard instances where GPU-based inprocessing provides limited benefit, the computational load shifts to the CDCL backend.

In contrast to GPU acceleration, MallobSat~\cite{schreiber2024mallobsat} focuses on large-scale distributed SAT solving using  high-performance computing (HPC) clusters. MallobSat overhauls the distributed solver HordeSat~\cite{balyo2015hordesat}, employing a fully CPU-based approach that orchestrates state-of-the-art backend solvers across hundreds of processors. Its core contribution lies in a communication-efficient clause sharing mechanism and a malleable design that adapts to fluctuating resources. However, this approach necessitates hundreds to thousands of CPU cores, which incurs significant costs and limits accessibility.

Recently, \textit{TurboSAT}~\cite{dai2025turbosat} proposes a hybrid scheme that exploits hardware heterogeneity, assigning parallel network training to the GPU and seeding the multi-threaded CPU solver with high-quality partial assignments. The authors represent the SAT problem as a binarized matrix-matrix multiplication layer and propose a differentiable objective function. 
While this approach achieves a 2$\times$ speedup over SOTA solver \textit{CaDiCaL}~\cite{biere2024cadical} under the PAR-2 metric,
it suffers from limitations in completeness and runtime efficiency. First, clause satisfaction is expressed as a matrix multiplication layer in which the discrete 
\textsc{and}/\textsc{or} semantics of clauses are approximated by linear scores. This score-driven optimization bypasses explicit logical inference, thereby limiting the solver’s efficiency to reason and converge. Second, the method remains incomplete and is unable to prove unsatisfiability (UNSAT).

%% file: 5-Conclusion.tex
\section{Conclusions}\label{results}

In this work, we introduce a complete and logically grounded SAT solving framework that integrates algebraic modeling with hybrid GPU-CPU execution in a dataless setting. By reformulating the SAT problem into a differentiable finite field algebraic form, our approach enables massively parallel clause evaluation and gradient-based optimization on GPUs. The GPU model produces a diverse set of partial assignments, which are distributed to multiple CPU threads for concurrent symbolic reasoning, achieving rapid convergence to complete solutions. 
Our experimental results demonstrate up to 8.41$\times$ speedup on \textit{Satisfiable} benchmarks and 1.29$\times$ on \textit{Unsatisfiable} benchmarks, far surpassing the state-of-the-art solvers. For perspective, while solver performance has improved by only about 2$\times$ over the past two decades, our method achieves comparable progress within a single framework.

Looking forward, we aim to further enhance scalability through algorithm–hardware co-design, improved memory utilization, and hierarchical batching to handle extremely large-scale CNFs (e.g., beyond 500 million clauses). Furthermore, the proposed differentiable algebraic formulation establishes a foundation for broader reasoning and EDA applications, such as SAT-based ATPG, Boolean synthesis, and technology mapping. Extending this framework to SMT solvers and differentiable formal verification, such as bounded model checking and logic equivalence checking, and exploring GPU-accelerated sequential reasoning represents promising directions for future research.

\section*{Acknowledgment}

This work was supported in part by the Defense Advanced Research Projects Agency (DARPA) under Contract No. HR001125C0058, and in part by the National Science Foundation (NSF) under Grants CCF-2403134 and CCF-2349670. The authors also thank Prof. Sharad Malik for valuable discussions and insightful feedback that contributed to this research.

%% file: 6-appendix.tex
\begin{appendices}

\subsection{Experiment Setup and Baselines}
All experiments were conducted on a server equipped with an AMD EYPC 9334 32-Core Processor and two NVIDIA A100 GPUs. The proposed framework was implemented in PyTorch, and evaluated on benchmarks from the SAT Competition 2024 dataset \cite{heule2024proceedings}, which consists of 179 satisfiable and 214 unsatisfiable instances. We compared the performance of our method against two state-of-the-art CDCL solvers, \textit{Kissat} and \textit{CaDiCaL}.

For training, the differentiable SAT formulation was configured with a fixed clause size of 3 after Tseitin normalization. This choice alleviates gradient vanishing in longer disjunctive clauses, where the multiplicative terms in Eq.~\ref{eq.eval} across many literals can lead to diminished gradient magnitudes during training.
Optimization was performed using the Adam optimizer with a learning rate of 0.5, a batch size of 3,000, and 10 training epochs. The temperature parameter in Eq.~\ref{eq.gumbel} was set to $\tau =1.0$. Each experiment was subject to a timeout limit of 5,000 seconds. For SAT instances, $N=100$ candidate solutions (Eq.~\ref{eq.pool}) were sampled and evaluated in parallel on CPU threads, where the top 0.05\% of high-confidence variables were used as partial assignments. For UNSAT instances, branching followed the procedure in Section~\ref{unsat_branching}, with $d=7$ selected variables forming $2^d=128$ instances, processed concurrently across 128 CPU threads. These thread counts (100 for SAT and 128 for UNSAT) were chosen to match the level of parallelism readily available on modern multi-core CPUs.

\subsection{Clause Normalization Example}



To demonstrate how the proposed differentiable pipeline operates in practice, we consider a Boolean formula in conjunctive normal form:
\begin{equation}
\varphi = (x_1 \vee \lnot x_2 \vee x_3 \vee x_4) \wedge (\lnot x_1 \vee x_3), \nonumber
\end{equation}
where $x_i$ denotes the $i$-th Boolean variable.
Applying Tseitin encoding to the 3-SAT form yields an equisatisfiable formula:

\begin{align}
\varphi' ={}&
(x_1 \vee \lnot x_2 \vee z_1) \wedge \nonumber\\
&(\lnot z_1 \vee \lnot x_3 \vee x_4) \wedge \nonumber\\
&(\lnot x_1 \vee x_3 \vee x_3),\nonumber
\label{eq:cnf}
\end{align}
where auxiliary variable $z1$ is introduced to decompose the longer clause into multiple clauses of uniform size. Following the Tseitin-transformed CNF $\varphi'$, each clause can be algebraically expressed as a polynomial over the Boolean variables using Eq.~\ref{eq.algebraic}. Then, the first clause $x_1 \vee \lnot x_2 \vee z_1$ is mapped to
\begin{equation}
    C_1 = x_1 + (1-x_2) + z_1 - x_1(1-x_2) - (1-x_2)z_1 -z_1 x_1 + x_1 (1-x_2) z_1,\nonumber
\end{equation}
which follows the general expansion rule for a three-term disjunction:
\[
a \vee b \vee c = a + b + c - ab - bc - ca + abc.
\]
Similarly, the remaining clauses are expressed as
\begin{align}
C_2 &= (1-z_1) + (1-x_3) + x_4  - (1-z_1)(1-x_3) - (1-x_3)x_4 - x_4(1-z_1) \nonumber \\
     &\quad + (1-z_1)(1-x_3)x_4, \nonumber\\
C_3 &= (1-x_1) + x_3 + x_3 
      - (1-x_1)x_3 - (1-x_1)x_3 - x_3x_3 
      + (1-x_1)x_3x_3. \nonumber
\end{align}

Suppose the Eq.~\ref{eq.gumbel} produces a two-dimensional probability vector, e.g., $y(x_1) = [0.7, \, 0.3]$, where the first entry corresponds to $x_1 = 0$ and the second to $x_1 = 1$. For simplicity, we only list the second entries of four variables $x_1$ to $x_4$ from  $\varphi'$ as $y = [0.7,\, 0.3,\, 0.2,\, 0.9]$. Auxiliary Tseitin variable $z_1$ is also parameterized and sampled in practice, but it is omitted here for clarity. Sampling from these distributions yields the hard assignment $\hat{x} =  [1,\, 0,\, 0,\, 1]$ according to Eq.~\ref{eq.argmax}, which then evaluates the clauses as $C_1 = 1, C_2 = 1$, and $C_3 = 0$ using Eq~\ref{eq.eval}. Thus, the loss becomes $\mathcal{L} = -(C_1 + C_2 + C_3) = -2$ by Eq.~\ref{eq.loss}.

\end{appendices}